\documentclass[10pt]{article}
\usepackage{amsmath}
\usepackage{amssymb}
\usepackage[english]{babel}
\usepackage[latin1]{inputenc}
\usepackage{graphicx}
\usepackage{eurosym}
\usepackage[round]{natbib}
\usepackage{subfig}
\usepackage{rotating}
\usepackage{multirow}
\usepackage{booktabs}
\usepackage{dcolumn}
\usepackage{longtable}
\usepackage{pdflscape}
\def\sym#1{\ifmmode^{#1}\else\(^{#1}\)\fi}

\newcommand\bss{\begin{sidewaystable}[htbp] \scriptsize \centering} 
\newcommand\ess{\end{sidewaystable}}
\newcommand\bs{\begin{table}[htbp] \tiny \centering}
\newcommand\es{\end{table}} 
\newcommand\notype[1]{\unskip}

\title{Keeping up with the e-Joneses: Do online social networks raise social comparisons?\footnote{The article was prepared within the framework of a subsidy granted to the HSE by the Government of the Russian Federation for the implementation of the Global Competitiveness Program. The usual disclaimers apply.}}
\author{Fabio Sabatini\footnote{Department of Economics and Law, Sapienza University of Rome, Italy, and LCSR National Research University Higher School of Economics, Russian Federation.} \footnote{Corresponding author. Postal address: Sapienza Universit\`{a} di Roma, Facolt\`{a} di Economia, via del Castro Laurenziano 9, 00161, Roma, Italy. E-mail: fabio.sabatini@uniroma1.it} and Francesco Sarracino\footnote{Institut national de la statistique et des \'{e}tudes \'{e}conomiques du Grand-Duch\'{e} du Luxembourg (STATEC), and LCSR National Research University Higher School of Economics,  Russian Federation. E-mail: Francesco.Sarracino@statec.etat.lu}}
\date{\today}

\begin{document}
\maketitle

\begin{abstract}
\begin{small}
Online social networks such as Facebook disclose an unprecedented volume of personal information amplifying the occasions for social comparisons.  We test the hypothesis that the use of social networking sites (SNS) increases people's dissatisfaction with their income. 
After addressing endogeneity issues, our results suggest that SNS users have a higher probability to compare their achievements with those of  others. This effect seems stronger than the one exerted by TV watching, it is particularly strong for younger people, and it affects men and women in a similar way. 

  
\end{small}
\end{abstract}

\newpage
\section{Introduction}
An old way of saying states: ``The neighbour's grass is always greener''.  People have the innate tendency to track their progress and assess their self-worth by comparing themselves to others. As a result,  satisfaction strongly depends on the possessions and achievements of individuals belonging to the so-called reference group. A reference group is the set of people that individuals use as a yardstick for evaluating themselves and their achievements. For example, they could be neighbours, colleagues, or people in the upper class -- in other words, all those   with whom people compare and to whom they wish to resemble.  

The action of comparing oneself with others in a reference group in order to evaluate or to enhance some aspects of the self is known as ``social comparison''. Such behaviour affects a variety of economic choices including consumption, investments in human capital, effort in the workplace, risk taking, and contribution to the provision of public goods just to name a few \citep{linde2012social, cfhs2014, gms2014}. In addition, social comparisons are a fundamental determinant of life satisfaction \citep{co1996, ferrer2005, daf2007}.

The possibility to compare oneself with others relies on the availability of information about the lives of others. A few economic studies have analysed how the ability of mass media to provide information on alternative lifestyles may in some way undermine self-esteem and well-being. \citet{bs2006b} and \citet{hs2013}  analysed the role of television. \citet{cs2010}  were the first to incidentally address the role of Internet access in a broader study about the intensity and direction of income comparisons. \citet{lohman2015}  systematically explored the effect of information and telecommunication technologies (ICT) with a specific focus on Internet access.

We add to this literature by testing the hypothesis that the use of social networking sites (SNS) such as Facebook raises people's tendency to compare themselves to others. We argue that online social networks, in fact, disclose an unprecedented volume of personal information that might be a powerful source of social comparisons. Our contribution bridges  two literatures. The first one was developed by economists who dealt with the roots of social comparisons \citep{stutzer2004, bs2006b, daf2007}. We add to these works by conducting the first study of the role of online social networks. The second literature encompasses psychological studies that considered the extent to which using Facebook prompts feelings of frustration and dissatisfaction in small and limited samples of users, generally composed of undergraduate students attending specific colleges \citep{vk2015, ly2015, tfd2015}.

Previous studies on social comparisons have found that, on average, individuals report comparing themselves to others about once per day \citep{wm1992}. There are several reasons to believe that this average may have increased with the increasing penetration of Facebook into users' daily life. First, users are likely to experience an extension of their reference group. Most users have Facebook ``friends'' who are actually past friends or distant acquaintances, whose information would not be as readily, if at all, accessible without Facebook. Several studies, in fact, have provided evidence that Facebook allows the crystallization of weak or latent ties that might otherwise remain ephemeral \citep[see for example][]{esl2007, ass2014}.

Moreover, because most Facebook users allow all ``friends'' unrestricted viewing of their profiles \citep{pyc2009}, individuals often have access to a large amount of information about even their most distant acquaintances. These posts may form the basis for numerous social comparisons.

Second, Facebook allows a more efficient access to information about ``friends'' compared to offline interactions. While individuals may not meet or engage in face-to-face conversations with even close friends on a daily basis, Facebook allows individuals to keep in touch with, and monitor the activities of, numerous friends multiple times a day, even when those friends are in different locations.

Most importantly, online social networks not only offer more frequent opportunities for comparison, but they also offer more opportunities for upward comparisons, i.e. towards those who look better off. This is due, in particular, to the prevalently positive nature of information that people choose to display on Facebook. Psychological studies have shown that Facebook tends to serve as an onslaught of idealized existences -- babies, engagement rings, graduations, new jobs, consumption of expensive goods and services such as cars and vacations -- that invites upward social comparisons at a rate that can make ``real life'' feel like just a grey routine. This evidence is supported by studies finding that heavy Facebook users were more likely to agree that others were ``happier'' and ``had better lives'' than individuals who used the site less frequently \citep{ce2012}. This evidence indicates that Facebook does leave users with a positively-skewed view of how others are doing, which can be a source of frustration and dissatisfaction with own life. 

Our analysis is based on the 2010, 2011, and 2012 waves of the Multipurpose Household Survey (MHS) provided by the Italian National Institute of Statistics (Istat). The survey includes individual level information about people's participation in SNS, a proxy of social comparisons, as well as a number of individual level control variables. To explore the association between the use of online social networks and social comparisons, we use a standard ordered probit regression model. Moreover, to account for possible endogeneity issues, we exploit the availability of fast Internet access across Italian regions -- which is related to the orographic features of the territory that determined the technological characteristics of the pre-existing voice telecommunication infrastructures.
Our results suggest that  SNS users     have a higher probability to compare their achievements with those of  others. This effect seems stronger than the one exerted by TV watching, it is particularly strong for younger people, and it affects men and women in a similar way.

The paper begins by providing the motivation of the study and briefly reviewing the literature (Section \ref{lit}). We then describe our data and empirical strategy in Section \ref{data}. In Section \ref{results} we present and discuss our results. Section \ref{conclu} concludes.

\section{Related literature}
\label{lit}

In economics, the study of the satisfaction and dissatisfaction driven by social comparisons can be traced back to the very origins of the concept of utility. As \citet{bentham1781} used it, utility refers to pleasure and pain, the ``sovereign masters'' that ``point out what we ought to do, as well as determine what we shall do'' \citep{kws1997}. Bentham explained how the pleasures and pains enjoyed and suffered by others are fundamental sources of human satisfaction and dissatisfaction. ``The pleasures of malevolence are the pleasures resulting from the view of any pain supposed to be suffered by the beings who may become the objects of malevolence''. ``The pains of malevolence are the pains resulting from the view of any pleasures supposed to be enjoyed by any beings who happen to be the objects of a man's displeasure'' \citep[][pp. 37-40]{bentham1781}. 

Marx explained the relative nature of utility in his early work on wage, labour, and capital: ``Our wants and pleasures have their origin in society; we therefore measure them in relation to society; we do not measure them in relation to the objects which serve for their gratification. Since they are of a social nature, they are of a relative nature'' \citep[][p. 45]{marx1847}. A few years later, \citet{veblen1899} introduced the concept of `conspicuous consumption', serving to impress other persons.

Nevertheless, the term social comparison was not introduced until the classic paper by \citet{Festinger1954} in which the author explained the role of social comparisons in evaluating own opinions and abilities: ``To the extent that objective, non-social, means are not available, people evaluate their opinions and abilities by comparison respectively with the opinions and abilities of others'' (p. 118). Festinger also implicitly introduced the concepts of downward and upward comparisons (which were later formalized by \citet{wills1981}) by arguing that  ``The tendency to compare oneself with other specific person decreases as the difference between his ability or opinions and one's own increases'' (p. 120).  For example, an undergraduate student in an average college does neither compare herself to inmates of an institution for feeble minded (which would be a ``downward comparison''), nor to colleagues attending a PhD programme in a top university (``upward comparison''). In fact, comparisons with so distant others would necessarily be inaccurate.

Yet, it was \citet{duesenberry1949}  the first one to empirically test the importance of relative income for utility. His results suggested that upward comparisons overcome downward comparisons in determining people's aspirations and satisfaction. Aspirations, in fact, tend to be above the level already reached. As a result, wealthier people impose a negative externality on poorer people, but not vice versa.

\citet{wm1992} explained that comparisons about performance (also called ``similar comparisons'') are more frequent between close friends. Upward and downward comparisons (also called ``dissimilar comparisons''), on the other hand, are more frequent in more distant relationships. This kind of comparison is more likely to occur in online social networks than in face-to-face interactions, since SNS like Facebook allow users to interact with -- or to silently observe the lifestyles of -- distant others such as friends of friends, distant acquaintances or public figures. 

\citet{bb1977}  suggested that close friends generally want to avoid upward and downward comparisons because they are concerned with the negative feelings that they might prompt. This may result in a particular delicacy in reporting about specific life events or achievements in face-to-face conversations with friends. For example, a happily married individual may want to use tact in talking about her marriage to a friend who has just divorced. SNS-mediated interactions, on the other hand, usually start with the unilateral sharing of information with an indistinct audience. In this context, individuals are less likely to be concerned with specific friends' feelings. The simplified forms of communication offered by SNS -- such as the acts of posting a ``status'' or sharing a photo -- offer less ways to adopt delicacy and tact in dealing with others. In addition, Facebook research has proved that most users tend to over-share the bright side of their lives -- e.g. consumption of vacations, culture, or expensive goods and services -- to impress others and to attain or maintain a given social status.

Even if face-to-face interaction provides many opportunities to witness the conspicuous consumption of friends, SNS-mediated interaction offers way more chances to acquire detailed information about friends', acquaintances', as well as distant or unknown others' lifestyles. For SNS users, it is virtually impossible to avoid seeing such information because of the very nature of the news feed, in which ``friends'' post their ``status'' on a regular basis.

The empirical literature has operationalized the concept of social comparisons through measures of income aspirations, relative deprivation, and dissatisfaction with income. The first tests of the role of  social comparisons suggested that individuals' income aspirations are influenced by face-to-face interactions. Using a cross-section of Swiss survey data, \citet{stutzer2004} showed that a higher income level in the community determines higher individual aspiration levels, and that the discrepancy between income and aspiration matters for well-being. The more an individual interacts with her neighbours, the more the income situation of the community where she lives matters in defining her aspirations. 

\citet{bs2006b} used World Values Survey (WVS) data to analyse the effect of television, an agent of consumption socialization, on income aspiration. Their results indicate that the effect of income on subjective well-being is significantly lower for heavy-TV viewers. Bruni and Stanca explain that ``by watching TV people are overwhelmed by images of people richer and wealthier than they are. This contributes to shifting up the benchmark for people's positional concerns: income and consumption levels are compared not only to those of their actual social reference group, but also to those of their virtual reference group, defined and constructed by television programs. As a consequence, television viewing makes people less satisfied with their income and wealth levels'' (2006, p. 213).

If television, a unidirectional mass medium that provides relatively limited information about the lives of others, affects income aspirations and viewers' satisfaction with their income, it stands to reason that online social networks, which allow interactive communication and provide an unprecedented volume of personal information, might affect income aspirations even more. 

Surprisingly, the role of social media has never been analyzed before in economics. Based on data drawn from the third wave of the European Social Survey, \citet{cs2010}  found that individuals with Internet access tend to attach greater importance to income comparisons. Using panel data from the European Union Statistics on Income and Living Conditions (EU-SILC), \citet{lohman2015}  found stated material aspirations to be significantly positively related to fast-Internet access. Lohmann also reported cross-sectional evidence from the WVS suggesting that people who regularly use the Internet as a source of information derive less satisfaction from their income. Due to lack of data, these authors could not assess how material aspirations relate to the use of online social networks.

A few psychological studies have assessed the possible effects of Facebook on users' self-esteem, feelings of deprivation, and subjective well-being. Based on an online survey of 736 college students recruited via email from a large Midwestern university, \citet{tfd2015}  found that the use of Facebook triggers feelings of envy, which expose users to the risk of depression. \citet{ly2015}  used the survey responses of 446 university students attending a Korean university to study the emotional effect of social comparisons occurring in a SNS environment. Their results suggest that a predominant activity in SNS is making social comparisons with public figures and that such comparisons trigger a range of emotional responses including envy and shame. Based on a survey administered to 231 young adults recruited by two students at the University of Amsterdam through their online social networks, \citet{vk2015}  found that Facebook use was related to a greater degree of negative social comparison, which was in turn related negatively to self-perceived social competence and physical attractiveness. The main limitation of this body of research resides in the use of small, delimited and biased samples, in most cases composed of self-selected groups of undergraduate students attending specific  colleges. Our study is the first to provide an assessment of the relationship between SNS use and social comparisons in a large and nationally representative sample. 

\section{Data and empirical strategy}
\label{data}

We used a pooled cross-section of data drawn from the 2010, 2011 and 2012 waves of the MHS provided by the Istat. The MHS sample includes approximately  20,000 households, corresponding to nearly 48,000 individuals, and it provides, among others, information on Italians' social interactions habits. 

To measure social comparisons, we use a proxy capturing people's dissatisfaction with their income. Income satisfaction is, indeed, strongly correlated with relative deprivation \citep{daf2007, Conchita} and several studies used financial dissatisfaction as a proxy of social comparisons \citep[see for example][]{Brockmann2009}. Seminal work in psychology theorized that dissatisfaction is tightly linked to social comparisons. For example, in their pioneer study on the attitudes of American soldiers during World War II, \citet{stouffer1949} found that soldiers' feelings of dissatisfaction with their own condition were less related to the actual degree of hardship they experienced than to the situation of the unit or group to which they compared themselves. In other words, dissatisfaction basically depends on social comparison. More recently, economic studies have ascertained that satisfaction with income and subjective well-being is driven by the gap between the individual's income and the incomes of all individuals richer than him \citep{co1996, bda2006, daf2007, Conchita}. 

Income satisfaction was measured by the MHS through responses to the question: ``How satisfied do you feel with your financial conditions?'', where possible responses were ``very satisfied'', ``fairly satisfied'', ``not much satisfied'' and ``not at all satisfied''. The scale of the answers has been reverted so that higher scores stand for more dissatisfaction. Income dissatisfaction among Italian macro-regions is highest in Southern Italy and in the islands: it varies between 2.27 and 2.57 in the North, 2.58 and 2.64 in the Center and 2.64 and 2.88 in the South (on a 4-points scale). People in North-Eastern regions tend to report the lowest income dissatisfaction. Additionally,   the average level of income dissatisfaction increased  almost everywhere between 2010 and 2012 (see figure \ref{fig0} and table \ref{averages} in the Appendix).

\begin{figure}[htbp]
\includegraphics[width=0.90\textwidth]{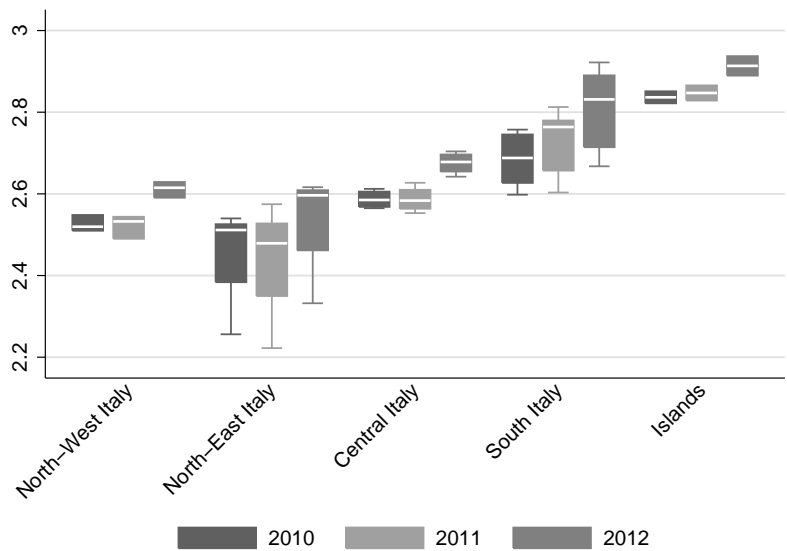}
\caption{Distribution of dissatisfaction with income  in Italy in 2010, 2011, and 2012.}
\label{fig0}
\end{figure}

The use of SNS was measured through a binary variable capturing respondents' use of online social networks such as Facebook and Twitter. Unfortunately, MHS data neither distinguish between Facebook and Twitter, nor contain information about the activities that users actually perform on these networks.

Since financial dissatisfaction, our dependent variable, is ordered in 4 categories, we adopt a standard ordered probit model. Formally, our baseline equation is as follows:  

\begin{equation}
\label{eq1}
\mbox{Income dissatisfaction}_{i} =
\begin{cases}
1 & \text{if~  $y_{i}$ $\leq$ 0,} \\
2 & \text{if~   $0~$  $<$ $y_{i}$ $\leq$ $c_1$,} \\
3 & \text{if~   $c_1$ $<$ $y_{i}$ $\leq$ $c_2$,} \\
4 & \text{if~   $c_{4}$ $<$ $y_{i}$.} \\
\end{cases}
\end{equation}

\begin{description}
	\item{where 0 $<$ $c_1$ $<$ $c_2$ $<$ $c_{3}$ $<$ $c_{4}$;}
	\item{the index $i$ stands for individuals;}
	\item{and $c_{1}$ - $c_{4}$ are unknown parameters to be estimated.}
	\item{$Y_{i} = \alpha + \beta_1 \cdot  fb_i + \boldsymbol{\theta} \cdot \mathbf{X_i} +  \varepsilon_{i} , \varepsilon_{i} \thicksim N(0,1)$}
\end{description}
\noindent
$Y_i$ is financial dissatisfaction, $fb_i$ is the use of SNS, $\boldsymbol{\theta}$ is a vector of parameters for the vector of control variables $\mathbf{X_i}$ and $\varepsilon_i$ is a vector of normally distributed errors with mean equal to zero and standard deviation equal to one. In all our regressions we account for possible heteroskedasticity of the errors using robust standard errors. The list of control variables includes: 
\begin{itemize}
\item{Fast Internet access, measured through the use of a broadband connection given by DSL or cable optical fibre. This control was included to test the hypothesis that broadband access raises material aspirations \citep{lohman2015} and to obtain hints about whether this relationship may be driven by the use of online social networks.}
\item{The frequency of meetings with friends, to check the possible relationship of face-to-face interactions with material aspirations \citep{stutzer2004}.}
\item{Age, gender, marital status, number of children in the household, education, and work status.}
\item{The number of minutes spent on watching TV per day. This control was included to further test the hypothesis that television raises material aspirations \citep{bs2006b}.}
\item{Macro-level controls including the real per capita GDP and the regional share of people  active in  volunteering activities. This variable was included to control for participation in associational activities that could provide people with opportunities of cooperative face-to-face interactions possibly capable of prompting social comparisons.}
\end{itemize}	

Descriptive statistics are reported in table \ref{desc}.

\begin{table}[htbp]
\caption{Descriptive statistics.}
\label{desc}

\begin{tabular}{lccccc} \hline
variable & mean & sd & min & max & obs \\ \hline
income dissatisfaction & 2.612 & 0.750 & 1 & 4 & 81499  \\
online networking & 0.460 & 0.498 & 0 & 1 & 38941 \\
women & 0.514 & 0.500 & 0 & 1 & 83092  \\
age & 49.45 & 18.25 & 18 & 90 & 83092  \\
age squared/100 & 27.79 & 18.99 & 3.240 & 81 & 83092  \\
minutes spent watching TV & 5.048 & 0.579 & 2.303 & 6.835 & 62602 \\
marital status & 1.954 & 0.842 & 1 & 4 & 83092  \\
educational status & 2.574 & 0.774 & 1 & 5 & 83092 \\
occupational status & 2.816 & 2.042 & 1 & 7 & 83092  \\
number of children & 1.023 & 1.009 & 0 & 7 & 83092  \\
modem & 0.107 & 0.309 & 0 & 1 & 48031  \\
DSL & 0.581 & 0.493 & 0 & 1 & 48031  \\
fiber & 0.0149 & 0.121 & 0 & 1 & 48031  \\
satellite & 0.0755 & 0.264 & 0 & 1 & 48031 \\
3G & 0.0244 & 0.154 & 0 & 1 & 48031  \\
USB & 0.178 & 0.382 & 0 & 1 & 48031  \\
mobile & 0.0193 & 0.138 & 0 & 1 & 48031  \\
fast internet connection & 0.596 & 0.491 & 0 & 1 & 48031  \\
real GDP per capita (thousands \euro 2005) & 22.95 & 5.730 & 14.58 & 30.77 & 83092 \\
regional share of volunteers & 0.104 & 0.0436 & 0.0537 & 0.231 & 83092  \\
region & -- & -- & 10 & 200 & 83092  \\
 year & -- & -- & 2010 & 2012 & 83092  \\ \hline
\end{tabular}
\end{table}

\subsection{Endogeneity issues}
\label{endog}
The coefficients from equation \ref{eq1} indicate the sign and magnitude of partial correlations among variables. However, we cannot discard the hypothesis that the use of SNS is endogenous to dissatisfaction with income. Personal characteristics such as, for example, a tendency for depression, may be correlated with both the use of SNS and our dependent variable. In addition, reverse causality might also arise. For example, dissatisfied people may want to use online social networks with the purpose of looking for new opportunities to improve their financial situation. 

To deal with such problems, we follow the strategy implemented in \citet{sabatini2014online, sabatini2015online}  and turn to instrumental variables estimates using a two stage least squares (2SLS) model where, in the first stage, we instrument the use of SNS.

We identified two  instruments: 
\begin{enumerate}
\item{The percentage of the population for whom a DSL connection was available in respondents' region of residence according to data provided by the Italian Ministry of Economic Development. DSL (digital subscriber line, originally digital subscriber loop) is a family of technologies that provides Internet access by transmitting digital data over the copper wires of a traditional local telephone network. Basically, it is a way to improve the speed of data transmission through old telephonic infrastructures. }
\item{A measure of the digital divide given by the percentage of the region's area that was not covered by optical fibre, elaborated from data provided by The Italian Observatory on Broadband. Optical fibre permits transmission over longer distances and at higher speed than DSL. }
\end{enumerate}
Both instrumental variables were measured in 2008, two years before the first wave of the Multipurpose Household Survey, which we employ in our study.

Even if DSL technology relies on the transmission of data over the user's copper old telephone line, the existence of a telephone infrastructure is just a necessary and not a sufficient condition for the availability of broadband. What matters is the so-called `local loop', i.e. the distance between final users' telephone line and the closest telecommunication exchange or `central office'. The longer the copper wire, the less bandwidth is available via this wire. If the distance is above a certain threshold (approximately 4.2 kilometres), then the band of the copper wires serving telephone communications cannot be wide enough to support a broadband connection \citep{campante2013}. When traditional telephone infrastructures were built, the length of copper wires was exogenously determined by the orographic features of the territory. If there were natural obstacles between users' telephone lines and the central office -- such as, for example, a hill or a lake -- then the length was likely to exceed the 4.2 kilometres threshold \citep{between2006, cs2008}. It seems therefore reasonable to assume that the 2008 level of regional DSL coverage cannot per se exert a direct influence on individuals' income dissatisfaction and that the influence of broadband coverage on the dependent variable actually depends on the extent to which accessing broadband allows Internet-mediated communication -- which in turn is a source of social comparisons.\footnote{In Appendix \ref{mappe}, we provide a map illustrating the orographic characteristics of the Italian territory and one showing the broadband coverage in 2007. The latter suggests that, in Italy, the most impervious territories are those with the worst broadband coverage.}

As for the second instrument, when the broadband connection cannot be implemented through pre-existing copper wires, it is necessary to turn to an optical fibre-based technology to provide fast-Internet. The possibility and the costs of installing this type of infrastructure, however, even more strongly rely on the exogenous characteristics of the natural environment. Differently from DSL, in fact, optical fibre entails the need to install new cables underground. This involves excavation works, which are expensive and generally delay or even prevent the provision of broadband in the area. As for DSL, orographic differences between regions must be considered as a `natural' cause of the variation in access to fibre across regions that is exogenous to people's  social interactions habits and cannot be driven by their preference for online networking. 
The tests of over-identifying restrictions  support the assumption of the orthogonality of the instruments.

For any given set of orographic characteristics of an area, the provision of broadband -- whether through DSL or optical fibre technology -- may also have been influenced by some socio-demographic factors that affected the expected commercial return on the provider's investment, such as population density, per capita income, the median level of education and the local endowments of social capital. These characteristics may correlate with our outcomes of interest in ways that could confound causal interpretation. To account for possible confounding effect, we included the regional level of per capita GDP, the regional share of volunteers, and a set of regional fixed effects to account for time invariant unobserved regional heterogeneity.

To instrument the use of SNS, we employ a 2SLS model. The first step   can be written as:

\begin{equation}
\label{eq2}
fb_i =  
\begin{cases}
0 & \text{if~  $y_{i}$ $\leq$ 0,} \\
1 & \text{if~  $y_{i}$ $>$ $0$.} \\
\end{cases}
\end{equation}
\noindent
where $fb_{i} = \pi_1 + \pi_2 \cdot z_1 + \pi_3 \cdot z_2  + \boldsymbol{\pi_4} \cdot \mathbf{X_i} + \nu_i, \nu_i\thicksim N(0,1)$ and $z_1$ and $z_2$ are the two above-mentioned instruments.  

The second step is as follows:

\begin{equation}
\label{eq3}
\mbox{Income dissatisfaction}_{i} =
\begin{cases}
1 & \text{if~  $y_{i}$ $\leq$ 0,} \\
2 & \text{if~   $0~$  $<$ $y_{i}$ $\leq$ $c_1$,} \\
3 & \text{if~   $c_1$ $<$ $y_{i}$ $\leq$ $c_2$,} \\
4 & \text{if~   $c_{4}$ $<$ $y_{i}$.} \\
\end{cases}
\end{equation}

\begin{description}
	\item{where 0 $<$ $c_1$ $<$ $c_2$ $<$ $c_{3}$ $<$ $c_{4}$;}
	\item{the index $i$ stands for individuals;}
	\item{$Y_{i} = \alpha + \beta_1 \cdot \hat{fb_i} + + \gamma_1 \cdot z_1 + \gamma_2 \cdot z_2 + \boldsymbol{\theta} \cdot \mathbf{X_i}  + \epsilon_{i} , \epsilon_{i} \thicksim N(0,1)$}
	\item{$\hat{fb_i}$ is the predicted probability of using SNS from the first step and $c_{1}$-$c_{4}$ are unknown parameters to be estimated.}
\end{description}
\noindent
As in model \ref{eq1}, $\boldsymbol{\theta}$ is a vector of parameters of the control variables $\mathbf{X}$; $\beta_1$ is the coefficient of SNS use; $\hat{fb_i}$ is the instrumented SNS use and $\epsilon_i$ is the error term.

To perform these estimates we used a multi-equation conditional mixed-process (CMP) estimator, as implemented by \citet{roodman2011}. This technique allows adopting a different specification of the model in each stage. In the first stage, where the dependent variable was the use of SNS, we used a probit model. In the second stage, the relation of SNS use with the indicator of financial dissatisfaction was estimated through an ordered probit model.

\section{Results}
\label{results}

Table \ref{tab1} presents the estimates from equation \ref{eq1}. Online networking is significantly and positively correlated with income dissatisfaction, thereby suggesting that people who use SNS tend to be more dissatisfied with their income, ceteris paribus. On the contrary, the higher is the frequency of meetings with friends, the lower is the respondent's income dissatisfaction. Together these two results suggests that face-to-face and web-mediated interactions might exert different effects on people's attitude to make social comparisons. This result might be related to the fact that, while SNS allow users to come into contact with distant others, such as acquaintances, past friends, or friends of friends, face-to-face interactions generally take place with close friends. The latter may prefer to avoid upward and downward comparisons for a matter of tact and delicacy, in that they are likely to be concerned with the negative feelings that might be associated with comparisons, as suggested by \citet{bb1977}.

Income dissatisfaction is also  significantly and positively associated with the amount of time spent  watching TV, which is consistent with previous findings documenting a major role of TV in raising people's material aspirations \citep{bs2006b}. As expected, broadband Internet is not significant,
though positively associated with income dissatisfaction. This suggests that the significant and positive relation between fast Internet use and measures of social comparisons found by \citet{cs2010} and \citet{lohman2015} may be due to the role of online social networks in providing personal information to their users.  All the other control variables have the expected signs. Income dissatisfaction is significantly higher for people with poor health and for people leaving in large households. On the contrary, married people and higher educated ones tend to compare less with  others. The coefficients of age and age squared document the existence of a U-shaped relationship between age and income dissatisfaction. Finally, we found that people living in richer regions tend to be less dissatisfied with their income, while we did not find any significant effect of the regional share of volunteers. These results hold after including regional fixed effects.

\bss
\def\sym#1{\ifmmode^{#1}\else\(^{#1}\)\fi}
\caption{Relationship between SNS and social comparisons \label{tab1}}
\resizebox{1.1\textwidth}{!}{
\begin{tabular}{l*{9}{D{.}{.}{-1}}}
\toprule
                &\multicolumn{1}{c}{(1)}         &         &\multicolumn{1}{c}{(2)}         &         &\multicolumn{1}{c}{(3)}         &         &\multicolumn{1}{c}{(4)}         &         \\
\midrule
women           &  -0.0409\sym{**} &  (-2.70)&  -0.0306\sym{*}  &  (-2.02)&  -0.0283\sym{*}  &  (-1.87)&  -0.0323\sym{*}  &  (-2.13)\\
age             &   0.0286\sym{***}&   (6.88)&   0.0279\sym{***}&   (6.69)&   0.0300\sym{***}&   (7.18)&   0.0301\sym{***}&   (7.17)\\
age squared/100 &  -0.0366\sym{***}&  (-7.77)&  -0.0360\sym{***}&  (-7.63)&  -0.0369\sym{***}&  (-7.83)&  -0.0371\sym{***}&  (-7.84)\\
good health     &    0.276         &   (1.36)&    0.255         &   (1.24)&    0.251         &   (1.22)&    0.249         &   (1.20)\\
neither good nor bad health&  -0.0627         &  (-0.32)&  -0.0706         &  (-0.36)&  -0.0754         &  (-0.38)&  -0.0713         &  (-0.36)\\
bad health      &   -0.357\sym{*}  &  (-1.83)&   -0.360\sym{*}  &  (-1.83)&   -0.363\sym{*}  &  (-1.85)&   -0.355\sym{*}  &  (-1.78)\\
very bad health &   -0.525\sym{**} &  (-2.68)&   -0.531\sym{**} &  (-2.69)&   -0.536\sym{**} &  (-2.71)&   -0.517\sym{**} &  (-2.58)\\
married         &   -0.208\sym{***}&  (-9.84)&   -0.225\sym{***}& (-10.56)&   -0.216\sym{***}& (-10.11)&   -0.216\sym{***}& (-10.10)\\
separated or divorced&   0.0220         &   (0.69)&  0.00945         &   (0.29)&   0.0110         &   (0.34)&   0.0196         &   (0.61)\\
widowed           &  -0.0824         &  (-1.23)&  -0.0996         &  (-1.48)&  -0.0941         &  (-1.39)&   -0.100         &  (-1.48)\\
primary&   -0.196         &  (-0.93)&   -0.231         &  (-1.09)&   -0.262         &  (-1.22)&   -0.251         &  (-1.16)\\
secondary&   -0.342         &  (-1.63)&   -0.397\sym{*}  &  (-1.88)&   -0.429\sym{*}  &  (-2.00)&   -0.407\sym{*}  &  (-1.88)\\
tertiary&   -0.532\sym{*}  &  (-2.53)&   -0.602\sym{**} &  (-2.84)&   -0.634\sym{**} &  (-2.95)&   -0.607\sym{**} &  (-2.80)\\
PhD &   -0.668\sym{**} &  (-3.06)&   -0.745\sym{***}&  (-3.40)&   -0.776\sym{***}&  (-3.48)&   -0.741\sym{***}&  (-3.31)\\
unemployed      &    0.784\sym{***}&  (29.38)&    0.708\sym{***}&  (26.08)&    0.708\sym{***}&  (26.09)&    0.694\sym{***}&  (25.45)\\
housewife       &    0.158\sym{***}&   (4.80)&    0.125\sym{***}&   (3.79)&    0.126\sym{***}&   (3.82)&    0.111\sym{***}&   (3.33)\\
student         &   0.0614\sym{*}  &   (1.96)&   0.0158         &   (0.50)&   0.0137         &   (0.44)&   0.0116         &   (0.37)\\
disabled        &    0.349\sym{**} &   (2.64)&    0.306\sym{*}  &   (2.32)&    0.309\sym{*}  &   (2.35)&    0.298\sym{*}  &   (2.25)\\
retired         &  0.00930         &   (0.25)&   0.0164         &   (0.45)&   0.0141         &   (0.38)&  0.00101         &   (0.03)\\
other work condition&    0.375\sym{***}&   (5.12)&    0.346\sym{***}&   (4.72)&    0.347\sym{***}&   (4.74)&    0.348\sym{***}&   (4.70)\\
number of children&   0.0640\sym{***}&   (8.05)&   0.0511\sym{***}&   (6.38)&   0.0525\sym{***}&   (6.56)&   0.0578\sym{***}&   (7.16)\\
frequency of meetings with friends&  -0.0304\sym{***}&  (-4.77)&  -0.0414\sym{***}&  (-6.44)&  -0.0442\sym{***}&  (-6.87)&  -0.0433\sym{***}&  (-6.70)\\
year 2011   &  0.00374         &   (0.25)&  0.00833         &   (0.56)&  0.00478         &   (0.32)&   0.0207         &   (1.23)\\
year 2012      &   0.0277         &   (1.05)&   0.0433         &   (1.63)&   0.0427         &   (1.61)&  -0.0322         &  (-0.78)\\
fast internet connection&   0.0122         &   (0.49)&   0.0406         &   (1.62)&   0.0297         &   (1.18)&   0.0186         &   (0.74)\\
mobile          &  -0.0502         &  (-0.84)&  -0.0276         &  (-0.46)&  -0.0446         &  (-0.74)&  -0.0669         &  (-1.11)\\
USB             &   0.0598\sym{*}  &   (2.07)&   0.0722\sym{*}  &   (2.49)&   0.0649\sym{*}  &   (2.23)&   0.0460         &   (1.57)\\
3G              &  -0.0802         &  (-1.61)&  -0.0564         &  (-1.13)&  -0.0698         &  (-1.39)&  -0.0846\sym{*}  &  (-1.69)\\
satellite       &  -0.0574\sym{*}  &  (-1.67)&  -0.0255         &  (-0.74)&  -0.0372         &  (-1.07)&  -0.0453         &  (-1.30)\\
real GDP per capita (thousands \euro 2005)&                  &         & -0.00818\sym{***}&  (-3.61)& -0.00859\sym{***}&  (-3.78)&   -0.149\sym{*}  &  (-2.33)\\
regional share of volunteers&                  &         &   -0.226\sym{***}&  (-7.07)&   -0.214\sym{***}&  (-6.69)&    4.130         &   (1.63)\\
online networking&                  &         &                  &         &    0.104\sym{***}&   (6.51)&    0.101\sym{***}&   (6.27)\\
minutes spent watching TV&                  &         &                  &         &                  &         &   0.0621\sym{***}&   (4.46)\\
\midrule

cut1        &   -2.119\sym{***}&  (-6.98)&   -1.963\sym{***}&  (-5.89)&   -1.937\sym{***}&  (-5.78)&   -14.39\sym{*}  &  (-2.06)\\

cut2        &   -0.108         &  (-0.36)&   0.0653         &   (0.20)&   0.0926         &   (0.28)&   -12.35\sym{*}  &  (-1.77)\\
 
cut3        &    1.014\sym{***}&   (3.34)&    1.194\sym{***}&   (3.59)&    1.223\sym{***}&   (3.65)&   -11.21         &  (-1.60)\\
\midrule
Observations    &    25379         &         &    25379         &         &    25379         &         &    25379         &         \\
Pseudo \(R^{2}\)&    0.043         &         &    0.048         &         &    0.049         &         &    0.053         &         \\
\bottomrule
\multicolumn{9}{l}{\footnotesize \textit{t} statistics in parentheses}\\
\multicolumn{9}{l}{\footnotesize \sym{*} \(p<0.1\), \sym{**} \(p<0.01\), \sym{***} \(p<0.001\)}\\
\multicolumn{9}{l}{\footnotesize Regional fixed effects omitted for brevity.}\\
\end{tabular}}
\ess

To address the endogeneity issues raised in Section \ref{endog}, we  turn to 2SLS estimates. Results are reported in table \ref{CMP}. Our two instruments are significantly and positively associated with the endogenous variable in the first stage. Additionally, the test of over-identifying restrictions confirms the validity of our instruments: as the coefficient of the Sargan test is not significant, we cannot reject the null hypothesis that  the over-identifying restrictions are valid. The coefficient of online networking is positive and significant. This supports the hypothesis that the use of social networking sites increases people's propensity to compare themselves to others. The remaining coefficients confirm results from the ordered probit model (see table \ref{tab1}).

\begin{table}[htbp]
\caption{Income dissatisfaction and use of facebook: IV estimates using CMP\label{tab3}}
\label{CMP}
\resizebox{13cm}{!}{
\begin{tabular}{l*{5}{D{.}{.}{-1}}}
\toprule
                & \multicolumn{2}{c}{\mbox{online networking } }            & \multicolumn{2}{c}{\mbox{income dissatisfaction}}         \\
\midrule

optic fiber (\%)&   0.0853\sym{*}  &   (1.76)&  -0.0767         &  (-1.04)\\
broadband coverage&   0.0595\sym{*}  &   (1.83)&  -0.0420         &  (-0.85)\\
women           &  -0.0760\sym{***}&  (-4.19)&  -0.0137         &  (-0.88)\\
age             &  -0.0507\sym{***}&  (-9.30)&   0.0426\sym{***}&   (8.83)\\
age squared/100 &   0.0158\sym{**} &   (2.48)&  -0.0416\sym{***}&  (-8.80)\\
good health     &   0.0920         &   (0.35)&    0.210         &   (1.04)\\
neither good nor bad health&    0.125         &   (0.49)&   -0.102         &  (-0.53)\\
bad health      &   0.0587         &   (0.23)&   -0.360\sym{*}  &  (-1.87)\\
very bad health &    0.101         &   (0.40)&   -0.524\sym{**} &  (-2.71)\\
married         &   -0.233\sym{***}&  (-9.47)&   -0.147\sym{***}&  (-5.53)\\
separated or divorced&  -0.0173         &  (-0.45)&   0.0292         &   (0.91)\\
widowed           &   -0.147\sym{*}  &  (-1.74)&  -0.0558         &  (-0.83)\\
primary &    1.346\sym{***}&   (3.84)&   -0.434\sym{*}  &  (-1.86)\\
secondary &    1.359\sym{***}&   (3.88)&   -0.589\sym{**} &  (-2.52)\\
tertiary &    1.378\sym{***}&   (3.93)&   -0.785\sym{***}&  (-3.36)\\
PhD &    1.301\sym{***}&   (3.66)&   -0.898\sym{***}&  (-3.76)\\
unemployed      &   0.0101         &   (0.33)&    0.660\sym{***}&  (21.64)\\
housewife       &  -0.0345         &  (-0.86)&    0.113\sym{***}&   (3.47)\\
student         &    0.123\sym{**} &   (3.25)& -0.00331         &  (-0.11)\\
disabled        &  -0.0996         &  (-0.66)&    0.304\sym{**} &   (2.41)\\
retired         &   0.0450         &   (0.89)&  -0.0113         &  (-0.31)\\
other work condition&  -0.0380         &  (-0.46)&    0.340\sym{***}&   (4.72)\\
number of children&  -0.0341\sym{***}&  (-3.52)&   0.0641\sym{***}&   (7.96)\\
frequency of meetings with friends&   0.0807\sym{***}&  (10.35)&  -0.0590\sym{***}&  (-8.39)\\
year 2011       &   0.0863\sym{***}&   (4.14)&  0.00114         &   (0.05)\\
year      2012 &   0.0847         &   (1.62)&  -0.0495         &  (-0.71)\\
minutes spent watching TV&   0.0624\sym{***}&   (3.85)&   0.0462\sym{**} &   (3.23)\\
fast internet connection&    0.328\sym{***}&  (10.35)&  -0.0501\sym{*}  &  (-1.72)\\
mobile          &    0.504\sym{***}&   (7.44)&   -0.170\sym{**} &  (-2.72)\\
USB             &    0.233\sym{***}&   (6.45)& -0.00330         &  (-0.11)\\
3G              &    0.397\sym{***}&   (6.61)&   -0.165\sym{**} &  (-3.14)\\
satellite       &    0.344\sym{***}&   (8.18)&   -0.115\sym{**} &  (-3.08)\\
real GDP per capita (thousands \euro 2005)&    0.147\sym{*}  &   (1.73)&   -0.174         &  (-1.34)\\
regional share of volunteers&   -2.308\sym{*}  &  (-1.84)&    2.333         &   (1.22)\\
online networking&                  &         &    0.754\sym{***}&   (5.57)\\
\midrule
cut1    &                  &     &   -21.54         &  (-1.17)        \\
cut2     &                  &    &   -19.59         &  (-1.06)       \\
cut3  &                  & &   -18.50         &  (-1.00)         \\
\midrule
Observations    &    25379         &         &    25379         &         \\
Wald chi$^2$       &            &         &         8774.6         &         \\
J\_stat          &                  &         &    1.548         &         \\
Prob $>$ J      &                  &         &    0.213         &         \\

\bottomrule
\multicolumn{5}{l}{\footnotesize \textit{z} statistics in parentheses}\\
\multicolumn{5}{l}{\footnotesize \sym{*} \(p<0.1\), \sym{**} \(p<0.05\), \sym{***} \(p<0.001\)}\\
\multicolumn{5}{l}{\footnotesize Regional fixed effects omitted for brevity.}\\
\end{tabular}}
\end{table}

\begin{figure}[htbp]
\caption{Marginal effects of the use of SNS on the probability of being very dissatisfied (upper charts) and very satisfied (lower charts) with own income situation. The charts on the left refer to people who do not use SNS, those on the right refer to the users of social networks.}
\label{marg4}
\centering {\bf (a) Income dissatisfaction}\\
\centering\includegraphics[scale=0.8]{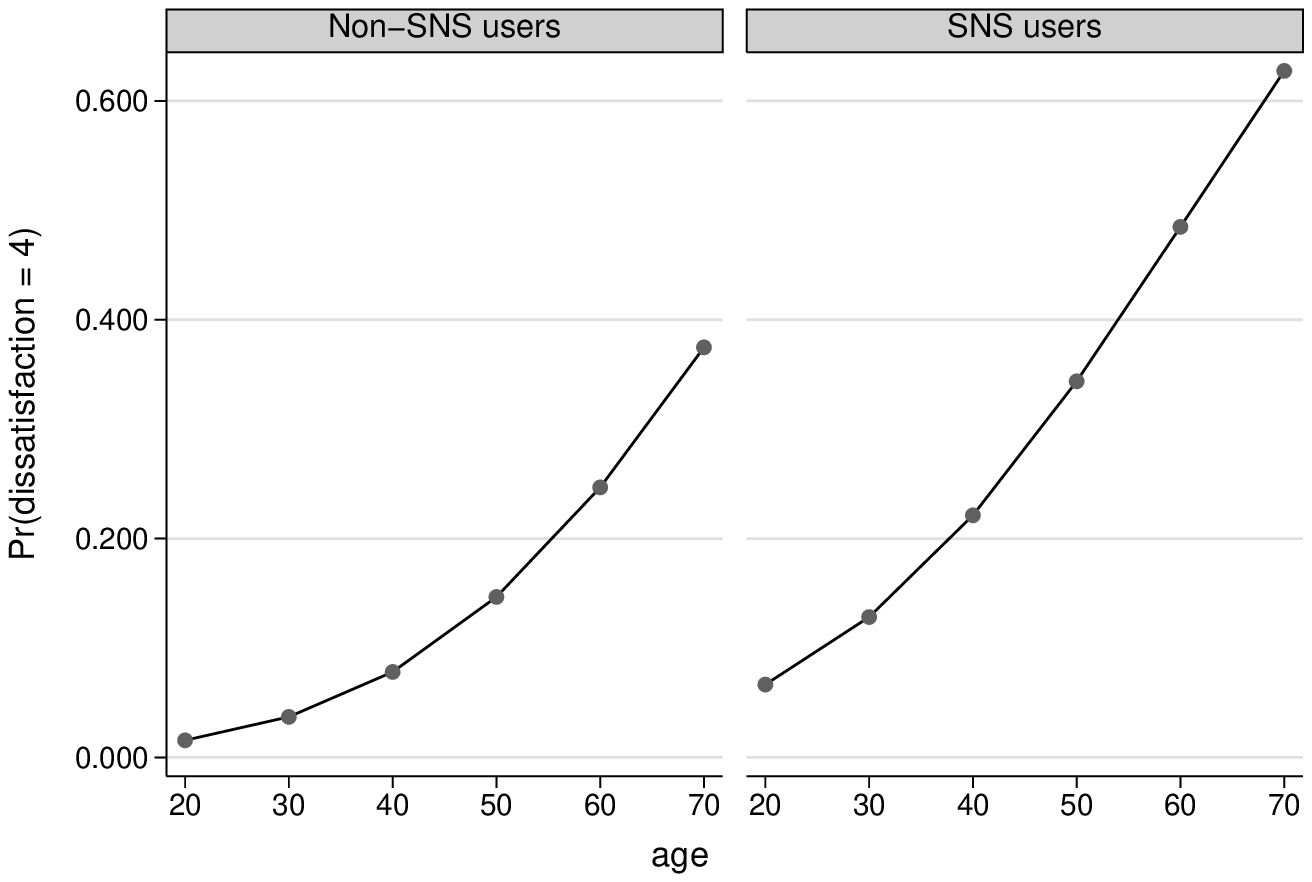}\\
\centering {\bf (b)  Income satisfaction} \\
\centering\includegraphics[scale=0.8]{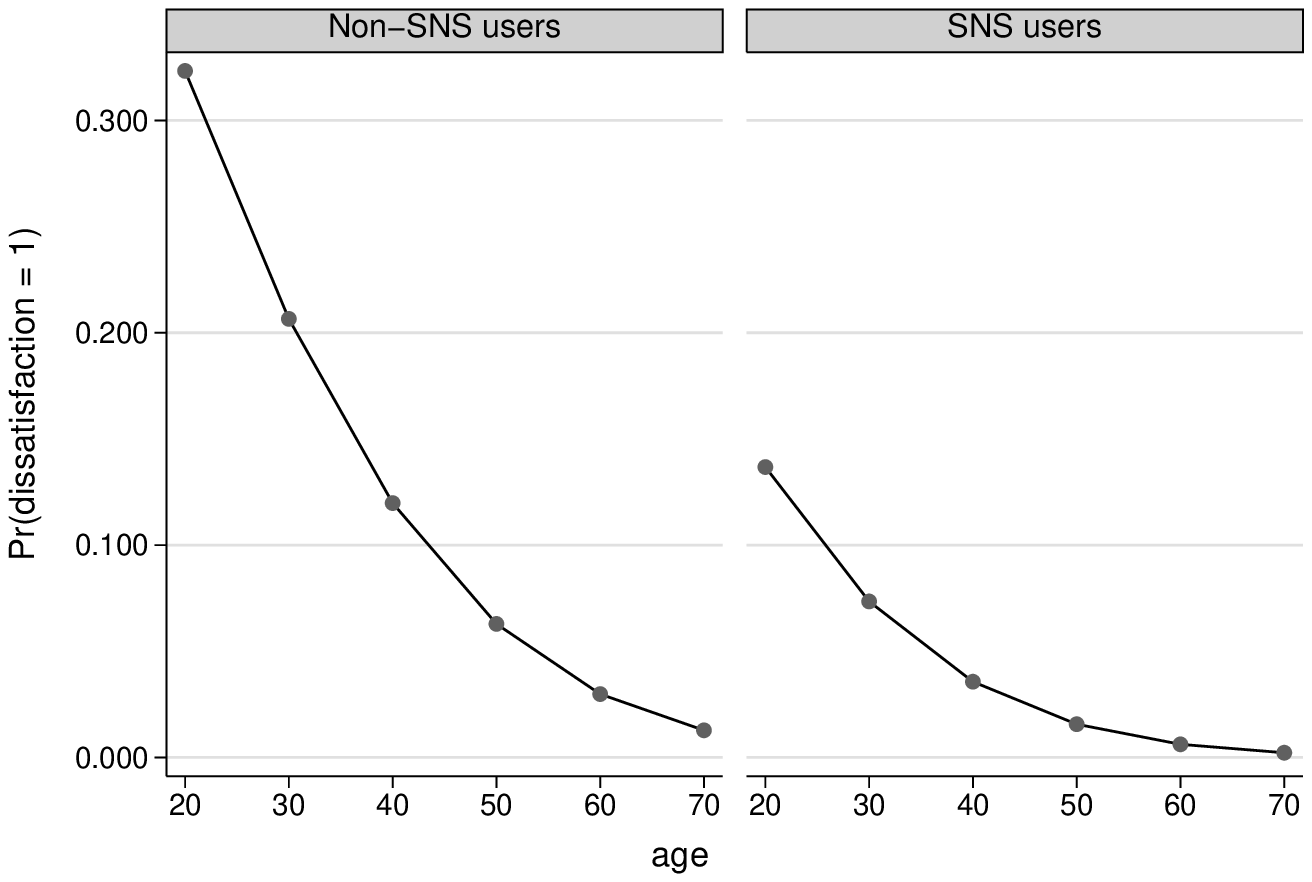}\\
\end{figure}

The upper right chart in figure \ref{marg4} shows the marginal effects of the use of social networking sites on income dissatisfaction  by age. The chart on the upper left allows to compare the same effects for people who do not use SNS. We notice a higher propensity to be dissatisfied with own income for older people. Yet, it is remarkable that, by each age, the marginal effect of using SNS on the probability of being very dissatisfied with own income   is higher than the marginal effect for those who do not use social networks. 

The charts in the lower part of figure \ref{marg4} provide the marginal effects of using and not using social networks on the probability of being very satisfied with own income   by age. The lines provide an  information  complementary to the one documented for income dissatisfaction: the probability of being very satisfied with own income   decreases with age and the marginal effects are stronger for social networks' users.

Table \ref{ame} reports the marginal effects of the use of SNS on the probability of being dissatisfied with own income after the instrumented estimates. The coefficients are increasingly positive and significant for the categories ``quite'' and ``a lot'' which suggests that the use of SNS increases the probability that people report to be at least quite dissatisfied with their income. Similarly, the second coefficient suggests that using SNS strongly reduces the probability to be ``a bit'' dissatisfied with own income.  The last coefficient, corresponding to the category ``not at all'', shows that using SNS slightly reduces the probability of declaring to be satisfied with own income: the coefficient is negative, but close to zero. In sum,  marginal effects document an increasingly positive effect  of using SNS on the probability of being very dissatisfied with own income.

\begin{table}[htbp]
\caption{Average marginal effects of the use of SNS on the probability of being  dissatisfied with own income.}
\label{ame}
\centering
\begin{tabular}{lccc}
\toprule
Pr(dissatisfaction) & dy/dx & Std. Err. & P-values \\
\midrule
not at all & -0.03\sym{***}  & 0.005  & 0.000  \\
a bit & -0.24\sym{***} & 0.042 & 0.000 \\ 
quite & 0.08\sym{***} & 0.008 & 0.000 \\
a lot & 0.19\sym{***} & 0.049 &  0.000 \\
\bottomrule
\end{tabular}
\end{table}

\begin{figure}[htbp]
\caption{Marginal effects of the use of SNS and TV watching on the probability to be satisfied  (left chart) and dissatisfied (right chart) with own income.}
\label{tvsns}
\centering\includegraphics[width=0.9\textwidth]{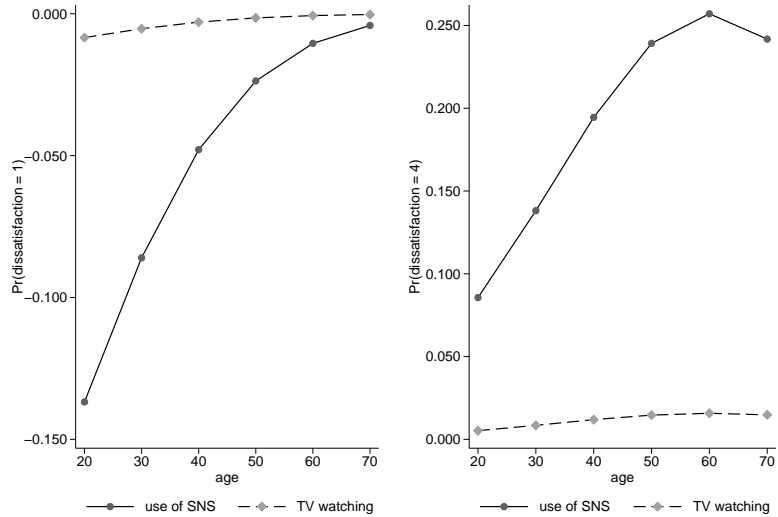}
\end{figure}

Overall, present results suggest that the use of SNS raises people's propensity to compare to others, making them less satisfied with their income.  In addition to supporting the claims that  television watching raises material aspirations, our results add to the literature by highlighting the  role of SNS and suggesting that previous findings about the role of broadband access in prompting social comparisons may actually be due to the use of SNS. Additionally, present results suggest that social networks play a major role in shaping people's tendency to compare to others. To illustrate this result, we compare the marginal effects of the use of SNS on income dissatisfaction with the marginal effects of TV watching (see figure \ref{tvsns}). The chart on the left in figure \ref{tvsns} shows that the marginal effect of TV watching on the probability of being satisfied with own income tends to zero when age increases. Yet, the coefficients are fairly small if compared to those related to the use of SNS. In the latter case we find strong negative effects of using SNS on the probability of being satisfied with own income: for people under the age of 40, using SNS reduces the probability of being satisfied with own income by more than 5\% and reaches the maximum of -13\% for people in their 20s. For people over 40 years old the same effect is smaller and gets close to the effect of TV watching for people in their 70s. 

The chart on the right of figure \ref{tvsns} shows the marginal effects of SNS use and TV watching on the probability of being very dissatisfied with own income. In this case we notice a small effect of TV watching which slightly increases with age. On the contrary, marginal effects suggest a strong role of SNS in income dissatisfaction which increases with age up to 60 years old. In other words, using SNS causes an increase of nearly 10\% in the probability of declaring to be dissatisfied with own income for people in their 20s. This percentage increases up to nearly 25\% for people in their 60s. This result suggests that income dissatisfaction of older people is more susceptible to the influence of SNS than the dissatisfaction of younger people. In sum, our findings support the hypothesis that SNS play a pivotal role in shaping people's comparisons to others.

\section{Conclusion}
\label{conclu}
Previous studies have highlighted the role of information in shaping positional concerns. In particular, the literature identified in television watching a vehicle of information about alternative lifestyles that stimulates social comparisons, which, in turn, can be an even more powerful cause of individuals' dissatisfaction with their life. 

Our results suggest that online social networks are a  powerful source of social comparisons. SNS, in fact, provide users with a volume of personal information that would have been unimaginable before the advent of Facebook, Twitter, and their companions. The power of online social networks in prompting comparisons is due to a number of factors. SNS allow users to monitor the activities and lifestyles not only of numerous friends but also of distant others, such as friends of friends, latent friends, or public figures, whose information would not be accessible without SNS. This information is strongly positively skewed because SNS users tend to over-share their positive life events and emotions and to allow unrestricted viewing of their information -- at least when it comes to positive one. As a result, the news feed of platforms like Facebook provides an onslaught of idealized existences that can boost upward comparisons to unprecedented levels. 

Even if people may compare themselves with an individual or group that they perceive as superior also as a means of self-improvement, in the hope that self-enhancement will occur, our results suggest that the social comparisons occurring in online social networks may cause frustration and dissatisfaction with income and material possessions, thereby lowering self-regard. This effect is strong and not homogeneous across age. Young people using SNS tend to have a lower probability to be satisfied with their income than older people, while the latter group tends to have higher probabilities of being dissatisfied with own income than younger people. Independently from age, present results document that using SNS is associated to a higher (lower) probability to be dissatisfied (satisfied) with own income. 

There are several reasons to treat our findings with prudence. The cross-sectional nature of the data employed in the analysis requires caution in advancing a causal interpretation of the estimates. The MHS lacks information about how much time users spend on SNS. It seems reasonable to argue that the more time people spend on platforms like Facebook, the more they assimilate news feed that provide updates, photos, and videos forming the bases for social comparisons. Most importantly, even if we are confident in the validity of our identification strategy, longitudinal data would help to more reliably identifying the effect of online social networks on social comparisons.

Despite these limitations, our study contributes to the literature by providing the first empirical investigation into the possible role of online social networks in social comparisons. The analysis also provides insights about the impact of social globalization -- a so far under investigated aspect of globalization -- on well-being. Our findings suggest that online social networks are an integral part of the social environment that embeds the economic action of individuals and play a vital role in determining people's satisfaction with their financial situation. Understanding how important economic decisions are made -- for example regarding consumption behavior and investments in human capital -- requires to deepen our knowledge of the impact of online social networks. The economic, social, and institutional context in which online social networks are used might also be of the greatest importance. Upward comparisons, in fact, might sort different effects depending on the conditions of the society to which SNS users belong. People in transition or fast growing countries may be more confident they will have the possibility to catch up with the standards of living of their online reference groups. In this case, the possibly negative effect of upward comparisons may be mitigated by the hope that an individual and a collective enhancement will occur. People living in developed countries with stagnating economy like Italy, on the other hand, may particularly suffer from the comparison processes related to social globalization. This suggests the need to carry out cross-country research about the possible effects of SNS.

\clearpage
\appendix

\section{Average levels of financial dissatisfaction in Italy.}
\begin{table}[htbp]
\caption{Average levels of dissatisfaction with the economic situation in Italy in 2010, 2011 and 2012.}
\label{averages}
\centering
\begin{tabular}{lccc} \hline
region & 2010 & 2011 & 2012 \\ \hline
Abruzzo & 2.627 & 2.603 & 2.714 \\
Basilicata & 2.657 & 2.748 & 2.820 \\
Calabria & 2.719 & 2.780 & 2.843 \\
Campania & 2.757 & 2.780 & 2.922 \\
Emilia Romagna & 2.512 & 2.477 & 2.616 \\
Friuli Venezia Giulia & 2.511 & 2.480 & 2.591 \\
Lazio & 2.599 & 2.627 & 2.704 \\
Liguria & 2.520 & 2.544 & 2.591 \\
Lombardia & 2.510 & 2.490 & 2.629 \\
Marche & 2.571 & 2.553 & 2.642 \\
Molise & 2.598 & 2.657 & 2.668 \\
Piemonte-Valle d'Aosta & 2.548 & 2.533 & 2.615 \\
Puglia & 2.746 & 2.813 & 2.890 \\
Sardegna & 2.822 & 2.829 & 2.890 \\
Sicilia & 2.851 & 2.866 & 2.937 \\
Toscana & 2.612 & 2.574 & 2.667 \\
Trentino Alto Adige & 2.256 & 2.222 & 2.332 \\
Umbria & 2.565 & 2.593 & 2.689 \\
 Veneto & 2.540 & 2.575 & 2.602 \\ \hline
\end{tabular}

\end{table}


\clearpage
\section{Orography and  broadband in Italy}
\label{mappe}
\begin{figure}[htbp]
\includegraphics[scale=0.55]{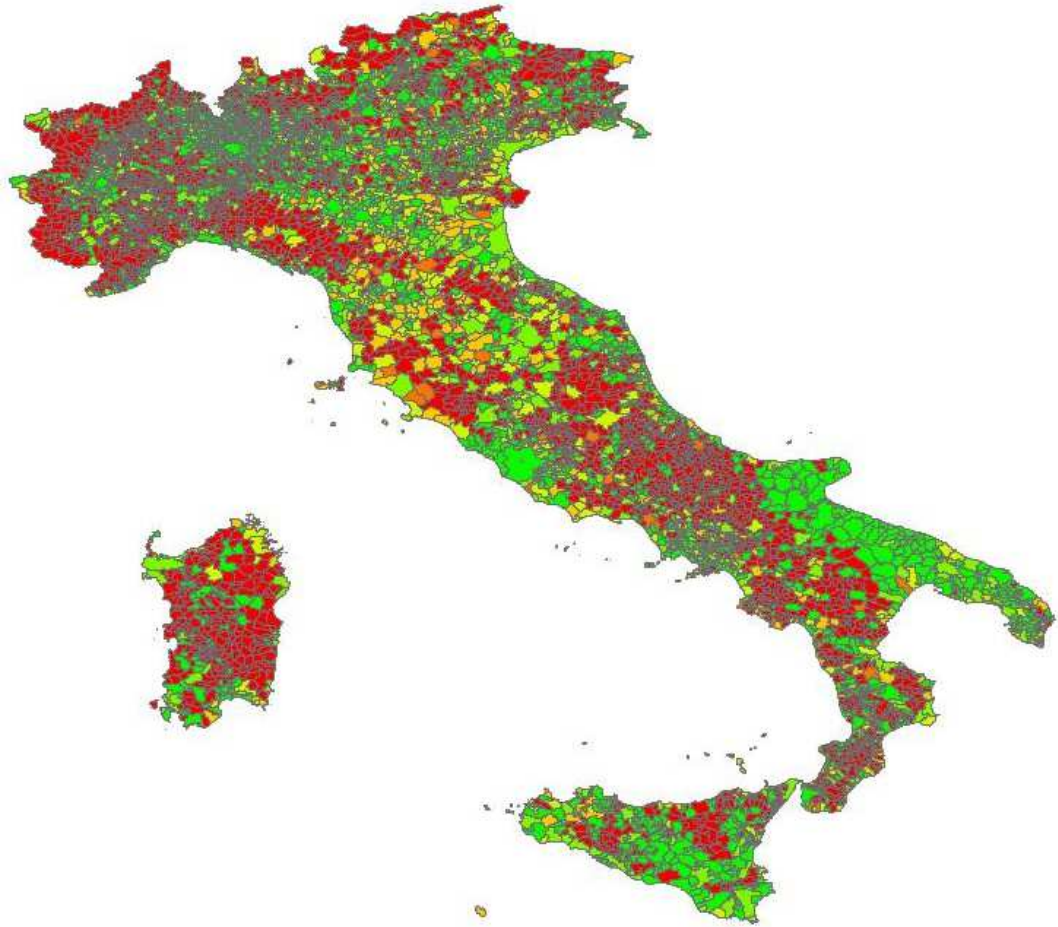}
\caption{Percentage of the population covered by broadband in Italy. \\
Source: Between (2006), p. 17. Darker areas are those with the worst coverage. Green areas have the best coverage.}
\label{mappabroad}
\end{figure}

\begin{figure}[htbp]
\centering\includegraphics[width=1.35\textwidth]{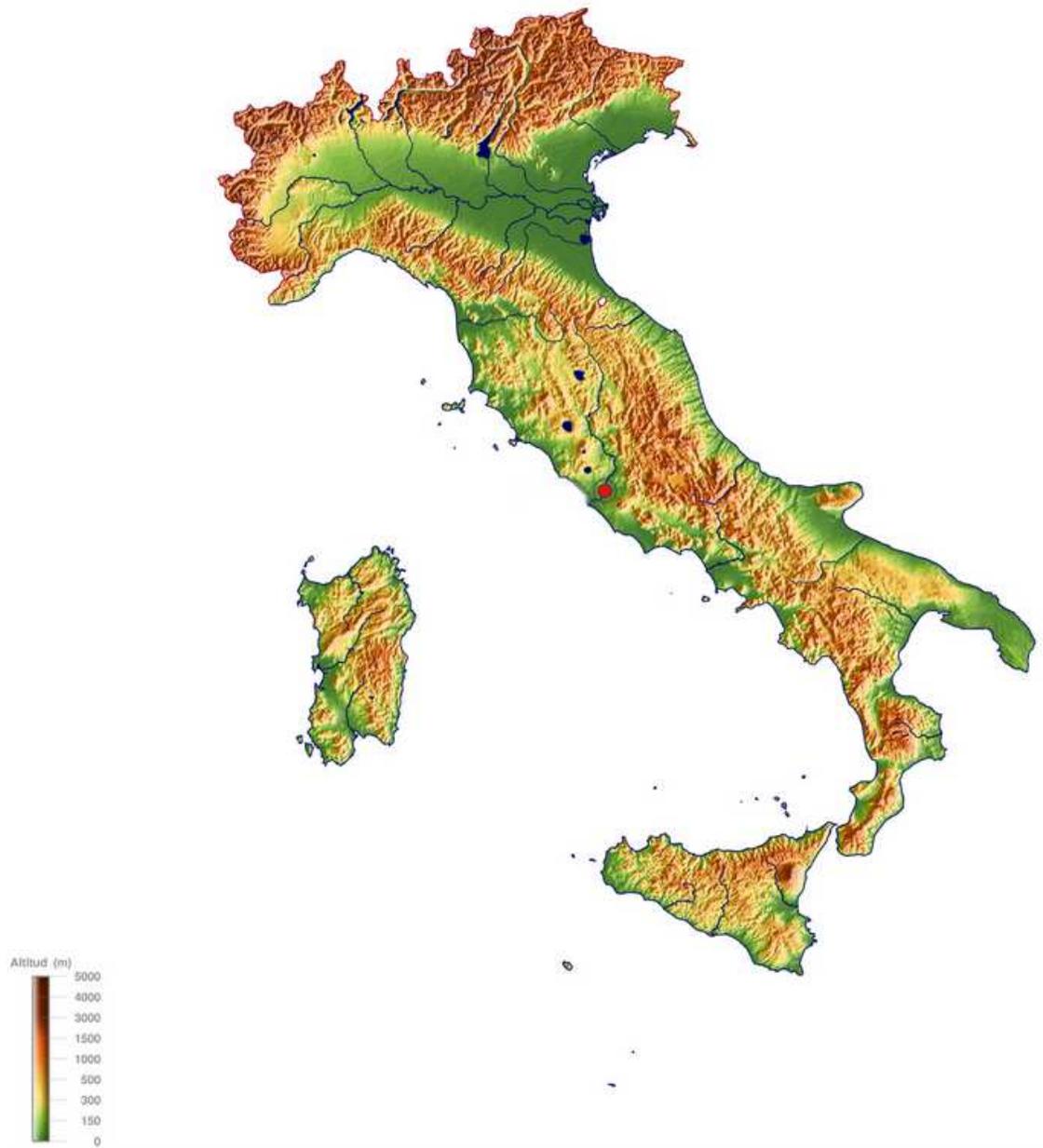}
\caption{Topographic map of Italy.}
\label{mappabroad}
\end{figure}

\clearpage

\end{document}